\documentclass[conference]{IEEEtran}
\usepackage[utf8]{inputenc}

\usepackage[T1]{fontenc}
\usepackage[normalem]{ulem}
\usepackage[english,french]{babel}

\usepackage{verbatim}
\usepackage{graphicx}
\usepackage{latexsym}

\usepackage{graphics}
\usepackage{epsfig}
\usepackage{amsmath}
\usepackage{rotating}
\usepackage{url}
\usepackage{fancyhdr}

\usepackage{tabularx}
\usepackage{array}
\usepackage{amsthm}
\usepackage{amsmath}
\usepackage{amssymb}
\usepackage{makeidx}
\usepackage{multirow}

\usepackage{multicol}
\usepackage{url}
\usepackage{textcomp}
\usepackage{cite}
\usepackage{float}
\usepackage{array}
\usepackage{caption}

\usepackage{etoolbox}
\usepackage{pifont} 
\usepackage{mdframed}

\theoremstyle{definition}
\newtheorem{thm}{Theorem}[section]

\newtheorem{pre}[thm]{Proposition}

\newtheorem{defini}[thm]{Definition}
\newtheorem{condi}[thm]{Condition}
\newtheorem{exemple}[thm]{Example}

\ifCLASSINFOpdf
\else
\fi
\usepackage{url}

\begin{document}
\selectlanguage{english}
\setlength{\columnseprule}{0pt}
\title{Secrecy by Witness-Functions under Equational Theories}

\author{\IEEEauthorblockN{Jaouhar Fattahi}
\IEEEauthorblockA{
 Department of Computer Sciences and\\ Software Engineering (LSI)\\Laval University, Quebec, QC, Canada\\
jaouhar.fattahi.1@ulaval.ca}

\and
\IEEEauthorblockN{Mohamed Mejri}
\IEEEauthorblockA{
Department of Computer Sciences\\ and Software Engineering (LSI)\\Laval University, Quebec, QC, Canada\\
 mohamed.mejri@ift.ulaval.ca
}

}


%



\maketitle

\begin{abstract}

In this paper, we use the witness-functions to analyze cryptographic protocols for secrecy under nonempty equational theories.  The witness-functions are safe metrics used to compute security. An analysis with a witness-function consists in making sure that the security of every atomic message does not decrease during its lifecycle in the protocol. The analysis gets more difficult under nonempty equational theories. Indeed, the intruder can take advantage of the algebraic properties of the cryptographic primitives to derive secrets. These properties arise from the use of mathematical functions, such as multiplication, addition, exclusive-or or modular exponentiation in the cryptosystems and the protocols. 
Here, we show how to use the witness-functions under nonempty equational theories and we run an analysis on the Needham-Schroeder-Lowe protocol under the cipher homomorphism. This analysis reveals that although this protocol is proved secure under the perfect encryption assumption, its security collapses under the homomorphic primitives. We show how the witness-functions help to illustrate an attack scenario on it and we propose an amended version to fix it. \\

\end{abstract}

\textit{\textbf{Keywords- Cryptographic protocols; Equational theories; Homomorphism; Secrecy.}}
\normalsize

%
\IEEEpeerreviewmaketitle


\section{INTRODUCTION}

In this paper, we use the witness-functions to statically analyze cryptographic protocols with respect to secrecy under nonempty equational theories. The Witness-Functions have been suggested by Fattahi et al. in \cite{WitnessArt1,WitnessArt2,2014arXiv1408.2774F,Fatt1407:Relaxed,IWSSS2014,ICNITJournal,suffCondWNIS} as metrics to attribute a safe value of security to each atomic message in the protocol. A protocol analysis  with a witness-function consists in following every atomic message defined in the protocol and making sure that its value of security does not fall down during its lifecycle. In this case, the protocol is said to be increasing -so correct- with respect to secrecy. The use of cryptographic primitives with algebraic properties compels the verifier to undertake special precautions when using these functions since the cryptographic primitives supply the intruder with new redoubtable capabilities. We organize this paper as follows:
\begin{itemize}
 \item First, we recall the theory of increasing protocols and we show that any protocol if proved increasing, using reliable metrics that meet few conditions, is correct with respect to secrecy;
\item then, we present the witness-functions as reliable metrics and we show how to use them under nonempty equational theories;
\item then, we run a formal analysis of the Needham-Schroeder-Lowe protocol and we show that although this protocol was proved correct under the perfect encryption assumption, it is no longer secure under the homomorphic primitives. We show that the witness-functions help to illustrate an attack scenario on it and we propose a corrected version of it based on hash functions;
\item finally, we compare our method to some related works. 
\end{itemize}

 \section*{ Notations}
Here, we give the notations used throughout this paper. 
\begin{itemize}
\item[+] We denote by ${\cal{C}}=\langle{\cal{M}},\xi,\models,{\cal{K}},{\cal{L}}^\sqsupseteq,\ulcorner.\urcorner\rangle$ the context of verification containing the relevant parameters for a protocols analysis.
\begin{itemize}
\item[$\bullet$] ${\cal{M}}$: is a set of messages built from the signature $\langle\cal{N}$,$\Sigma\rangle$ where ${\cal{N}}$ is a set of atomic names (nonces, keys, principals, etc.) and $\Sigma$ is a set of operators ($\mathcal{E}$:\!: encryption\index{Encryption}, $\mathcal{D}$:\!: decryption\index{Décryption}, $pair$:\!: pairing (denoted by "." here), etc.). i.e. ${\cal{M}}=T_{\langle{\cal{N}},\Sigma\rangle}({\cal{X}})$. We use $\Gamma$ to denote the set of substitution from $ {\cal{X}}$ to ${\cal{M}}$.
We denote by $\cal{A}$ the set of atomic messages in ${\cal{M}},$ by ${\cal{A}}(m)$ the set of atomic messages in $m$, by ${\cal{I}}$ the set of agents (principals) in the protocol and by $I$ the intruder. We denote by $k{^{-1}}$ the reverse  key of $k$ and we assume that $({k^{-1}})^{-1}=k$.
\item[$\bullet$] $\xi$: is the equational theory~\cite{Pigozzi1979117,cortier9900,cortier1971,cortier9901} that describes the algebraic properties of the operators in $\Sigma$ by equations. For example, the homomorphic property is described by $\{m.m'\}_{k}=\{m\}_{k}.\{m'\}_{k}$ and the modular exponentiation property is described by $\{\{m\}_{k}\}_{k'}=\{\{m\}_{k'}\}_{k}$. Two messages $m$ and $m'$ that are equivalent under the equational theory $\xi$ are denoted by $m=_\xi m'$.
\item[$\bullet$] $\models$: is the inference system of the intruder under the equational theory. Let $M$ be a set of messages and $m$ a single message. $M$ $\models$ $m$ expresses that the intruder is able to infer $m$ from $M$ using his capabilities. We extend that notation to valid traces as follows: $\rho$ $\models$ $m$ expresses the fact that the intruder can derive $m$ from the trace $\rho$.
\item[$\bullet$] ${\cal{K}}$ : is a function from ${\cal{I}}$ to ${\cal{M}}$, that attributes to any agent a set of atomic messages describing her initial knowledge. $K_{{\cal{C}}}(I)$ denotes the initial knowledge of the intruder( or simply $K(I)$ where the context is evident).
\item[$\bullet$] ${\cal{L}}^\sqsupseteq$ : is the security lattice $({\cal{L}},\sqsupseteq, \sqcup, \sqcap, \bot,\top)$ used to attribute a security level to a message. 
An example of a concrete lattice is $ (2^{\cal{I}},\subseteq,\cap,\cup,\cal{I}, \emptyset)$. We use this latter in this paper. 
\item[$\bullet$] $\ulcorner .\urcorner$ : is a partial function that attributes a value of security (type) to a message in ${\cal{M}}$. Let $M$ be a set of messages and $m$ be a sigle message. We express by $\ulcorner M \urcorner \sqsupseteq \ulcorner m \urcorner$ the fact
$\exists m' \in M. \ulcorner m' \urcorner \sqsupseteq \ulcorner m \urcorner$
\end{itemize}
\item[+] Let $p$ be a protocol, we denote by $R_G(p)$ the set of the generalized roles of $p$. A generalized role is a  protocol abstraction where the emphasis is put on a specific agent and where every unknown message by that agent and on which he cannot perform any verification is replaced by a variable. Further details on the role-based specification are available in~\cite{Debbabi11,Debbabi22,Debbabi33}.
We denote by ${\cal{M}}_p^{\cal{G}}$ the set of messages (ground terms and terms with variables) generated by $R_G(p)$, by ${\cal{M}}_p$ the set of messages that are ground terms generated by substitution in the messages of ${\cal{M}}_p^{\cal{G}}$. We denote by $R^-$ (respectively $R^+$) the set of received messages (respectively sent messages) by an agent in the role $R$. By convention, we reserve the uppercases for sets or sequences and lowercases for single items. For instance, $M$ denotes a set of messages, $m$ a message, $R$ a role consisting of a sequence of steps, $r$ a single step and $R.r$ the role ending by the single step $r$.
\item[+] A valid trace is a ground term obtained by substituting a non ground term in the generalized roles. We denote by $ [\![p]\!]$ the infinite set of valid traces.
\item[+] We assume that the intruder has the full-control of the net as described in the Dolev-Yao model~\cite{DolevY1}. We suppose no limitation neither on the length of messages nor on the number of interleaving sessions.
\end{itemize}

\section{About the Correctness of Increasing Protocols}\label{sectionPreuveThFond}

Hereafter, we recall a major result of the increasing protocols\cite{WitnessArt1,Fatt1407:Relaxed}: an increasing protocol is correct with respect to secrecy. For that, we need reliable metrics (functions) to estimate the security of the atomic messages of a protocol. To be reliable, a metric should meet few conditions. Here, we give these conditions and we substantiate the correctness of increasing protocols.

\subsection{Reliable Functions}
\begin{defini}{(Well-formed Function)}\label{bienforme}
{
Let ${F}$ be a function and ${\mathcal{C}}$ be a context of verification. ${F}$ is ${\mathcal{C}}$-well-formed iff:
$\forall M,M_1,M_2 \subseteq {\mathcal{M}}, \forall \alpha \in {\mathcal{A}}({\mathcal{M}}) \mbox{:}$
$
\left\{
\begin{array}{ll}
{F}(\alpha,\{\alpha\})= \bot; & \\
{F}(\alpha, {M}_1 \cup {M}_2)= {F}(\alpha, {M}_1)\sqcap{F}(\alpha,{M}_2); & \\
{F}(\alpha,{M}) =\top, \mbox{ if } \alpha \notin {\mathcal{A}}({M}). & \\
\end{array}
\right.
$
}
\end{defini}

A well-formed function ${F}$ should return the bottom value in the lattice for an atom $\alpha$ that appears in clear in $M$ to express the fact that is exposed to everybody in $M$. It should return for it in the union of two sets, the minimum of the two values calculated in each set alone. It returns the top value in the lattice for any atom $\alpha$ that does appear in $M$ to express the fact that none could derive it from $M$.

\begin{defini}{(Full-Invariant-by-Intruder Function)}\label{spi}
{
Let ${F}$ be a function and ${\mathcal{C}}$ be a context of verification. 
${F}$ is ${\mathcal{C}}$-full-invariant-by-intruder iff:\\
$
\forall {M} \subseteq {\mathcal{M}}, m\in {\mathcal{M}}. {M} \models_{\mathcal{C}} m \Rightarrow \forall \alpha \in {\mathcal{A}}(m). ({F}(\alpha,m) \sqsupseteq{F}(\alpha,{M})) \vee (\ulcorner K(I) \urcorner \sqsupseteq \ulcorner \alpha \urcorner).
$
}
\end{defini}

An full-invariant-by-intruder function ${F}$ is such that, when it affects a security value to an atom $\alpha$ in a set of messages $M$ the intruder can never deduce from $M$, using his capabilities, another message $m$ in which this value decreases (i.e. ${F}(\alpha,m) \not \sqsupseteq{F}(\alpha,{M})$), except when $\alpha$ is deliberately destined to the intruder (i.e. $\ulcorner K(I) \urcorner \sqsupseteq \ulcorner \alpha \urcorner$).

\begin{defini}{(Reliable Function)}
{
Let ${F}$ be a function and ${\mathcal{C}}$ be a context of verification.
\[{F} \mbox { is }{\mathcal{C}}\mbox{-reliable } \mbox{ iff } \left\{
\begin{array}{ll}
{F} \mbox{ is } {\mathcal{C}}\mbox{-well-formed} &
\\
{F} \mbox{ is } {\mathcal{C}}\mbox{-full-invariant-by-intruder}& 
\end{array}
\right.
\]
}
\end{defini} 
A reliable function ${F}$ is well-formed and full-invariant-by-intruder.

\begin{defini}{(${F}$-Increasing Protocol)}\label{ProAbsCroi}
{
Let ${F}$ be a function, ${\mathcal{C}}$ be a context of verification and $p$ be a protocol.\\
$p$ is ${F}$-increasing in ${\mathcal{C}}$ iff:\\
$\forall R.r \in R_G(p),\forall \sigma \in \Gamma: {\mathcal{X}} \rightarrow {\mathcal{M}}_p \mbox{ we have: }$
\[
\forall \alpha \in {\mathcal{A}}({\mathcal{M}}).{F}(\alpha, r^+\sigma)\sqsupseteq \ulcorner \alpha \urcorner \sqcap{F}(\alpha, R^-\sigma)
\]
}
\end{defini}

An ${F}$-increasing protocol generates permanently strings such that every atomic message in has always a security value, computed by ${F}$, higher in the sent message (i.e. in $ r^+\sigma$) than it was in the received messages (i.e. in $R^-\sigma$).

\begin{thm}{(Correctness of Increasing Protocols)}\label{mainTh}
{
Let ${F}$ be a ${\mathcal{C}}$-reliable Function and $p$ an ${F}$-increasing protocol.
\begin{center}
$p$ is correct with respect to secrecy.
\end{center}
}
\end{thm}

Theorem \ref{mainTh} states that a protocol is correct with respect to secrecy when it is increasing using a reliable metric $F$ to compute security. Hence, if the intruder manages to obtain a secret $\alpha$, then its value computed by $F$ is the bottom value in the lattice because $F$ is well-formed. This could not arise because of the protocol rules because the protocol is increasing on $F$ unless the value of security of $\alpha$ is the bottom from the beginning. In this case, $\alpha$ is not  a secret. That could not arise using the capabilities of the intruder neither since $F$ is full-invariant-by-intruder. Hence, the secret cannot be revealed. For further details on the proof, please see\cite{Fatt1407:Relaxed}.

\section{Building Reliable Functions under Equational Theories}\label{sectionFonctionsetSelections}\index{Fonction d'interprétation}

\subsection{Reliable Selections Under the Perfect Encryption Assumption}
In~\cite{WitnessArt1} we propose an abstract class  of reliable selections under the perfect encryption assumption that we denote by $S_{Gen}^{EK}$. Each selection $S$ in $S_{Gen}^{EK}$ should return for an atom $\alpha$ in a message $m$:

\begin{enumerate}
\item if $\alpha$ is encrypted by a key $k$ such that $k$ is the most external key satisfing the condition $\ulcorner k^{-1} \urcorner \sqsupseteq \ulcorner \alpha \urcorner$ (we call it the external protective key), a subset among $k^{-1}$ and the atoms that travel with $\alpha$ under the same protection by $k$ ($\alpha$ itself is not selected);
\item for two messages joined by a function $f$ in $\Sigma$ such that $f$ is not an encryption by the external protective key (e.g. pair), the union of the two subselections performed in each message separately.
\item if $\alpha$ does not have a protective key in $m$, the bottom value in the lattice (all the atoms);
\item if $\alpha$ does not appear in $m$, the top value in the lattice (the empty set);
\end{enumerate}


From the abstract class  $S_{Gen}^{EK}$, we propose three usefull selections:
\begin{enumerate}
\item the selection $S_{MAX}^{EK}:$ returns for an atom $\alpha$ in a message $m$ encrypted by the external protective key $k$, all the principal identities under the same protection by $k$, in addition to $k^{-1}$;
\item the selection $S_{EK}^{EK}:$ returns for an atom $\alpha$ in a message $m$ encrypted by the external protective key $k$, only the key $k^{-1}$;
\item the selection $S_{N}^{EK}:$ returns for an atom $\alpha$ in a message $m$ encrypted by the external protective key $k$, all the principal identities under the same protection by $k$;
\end{enumerate}

\subsection{Reliable Selections Under Equational Theories}

In nonempty equational theories~\cite{Pigozzi1979117,cortier9900,cortier1971,cortier9901}, cryptographic primitives have algebraic properties that arise from the use of mathematical functions like multiplication, addition, exclusive-or or modular exponentiation in cryptosystems and protocols. In Example\ref{eqthexp} we provide some of these algebraic properties. 

\begin{exemple}(Some Algebraic Properties)\label{eqthexp}
$ $\\
\begin{itemize}
\item Homomorphism: is the property that leads to have an equivalence between the two terms $\{m.m'\}_k$ and $\{m\}_k.\{m'\}_k$.  That is the case of the RSA public key cryptosystems, the ElGamal cryptosystem, the Brakerski-Gentry-Vaikuntanathan cryptosystem, the NTRU-based cryptosystem, the Gentry-Sahai-Waters cryptosystem, the Goldwasser–Micali cryptosystem, etc; 
\item Modular exponentiation: is the property that leads to have an equivalence between the two terms $\{\{m\}_k\}_{k'}$ and $\{\{m\}_{k'}\}_k$. This is the case of the Diffie-Hellman key agreement protocol;
\item XOR cipher: in many encryption algorithms, a plaintext is encrypted by applying the bitwise XOR operator to each character using some key $k$. To decrypt the output, applying the XOR function over with the key will cancel out the cipher. The XOR operator is vulnerable to a known attack since plaintext XOR ciphertext = $k$;
\item Etc.
\end{itemize}
\end{exemple}

These properties endow the intruder with additional capabilities to manipulate the protocol. 

\begin{condi}(Normal form with the smallest selection)\label{cond1}
Let $S$ be a selection of the class $S_{Gen}^{EK}$ and ${\cal{C}}$ be a context of verification. Let's have a rewriting system $\rightarrow_{\xi}$ such that $\forall m \in {\cal{M}},\forall \alpha \in {\cal{A}}(m) \wedge \alpha \not \in Clear(m),$ we have: $$ \forall l \rightarrow r \in \rightarrow_{\xi},S(\alpha,r) \subseteq S(\alpha,l)$$
We denote by $m_\Downarrow$ the normal form of $m$ in $\rightarrow_{\xi}$.
\end{condi}

 The condition on the rewriting system is introduced to make sure that the selection in the normal form is the smallest among all forms of a given message. This prevents the selection $S$ to select atoms that might be inserted maliciously by the intruder by manipulating the equational theory. Hence, we are sure that all selected atoms by $S$ are honest and do not come by an intruder manipulation of the message. We assume that the equational theory in the context of verification allows always the extraction of a convergent rewriting system that meets Condition \ref{cond1}. This is the case with the most of equational theories used in the literature~\cite{cortier9900,cortier1971,cortier9901}.

\begin{exemple}
Let $m=\{\alpha.C\}_{k_{ab}}$ be a message. Let us have a context of verification that includes the homomorphic cryptography (i.e. $\{\alpha.C\}_{k_{ab}}=\{\alpha\}_{k_{ab}}.\{C\}_{k_{ab}}$). In the form $\{\alpha.C\}_{k_{ab}}$, the selection $S(\alpha,\{\alpha.C\}_{k_{ab}})$ can select $C$, but in the form $\{\alpha\}_{k_{ab}}.\{C\}_{k_{ab}}$, the selection $S(\alpha,\{\alpha\}_{k_{ab}}.\{C\}_{k_{ab}})$ cannot. We orient so the rewriting system so that it returns the form $\{\alpha\}_{k_{ab}}.\{C\}_{k_{ab}}$ that is the normal form we choose.
\end{exemple}

\subsection{From Selections to Reliable Functions Under Equational Theories}
Having defined the selections above, we transform them now to security values. For that, we compose any selection $S$ in $S_{Gen}^{EK}$ with a suitable morphism $\psi$ and this composition leads to a reliable function $F=\psi \circ S$. 
We define the morphism as follows: 
\begin{enumerate}
\item it returns for a principal, its identity;
\item it returns for a key $k^{-1}$, if selected, the set of principals that know it in the context of verification.
\end{enumerate}
We denote by $F_{MAX}^{EK}, F_{EK}^{EK}$ and $F_{N}^{EK}$ respectively the functions resulting from the compositions $\psi \circ S_{MAX}^{EK}, \psi \circ S_{EK}^{EK}$ and $\psi \circ S_{N}^{EK}$ and we prove that these functions are ${\mathcal{C}}$-reliable. The main idea of the proof is that the selection for any secret $\alpha$ in a message is carried out  in an invariant zone (piece of message) that could not be augmented by the intruder using the equational theory seeing that the rewriting system is oriented in such way that the used form of a message is the smallest and contains always honest atoms only. This zone is in addition protected by a protective key $k$ that meets the condition  $\ulcorner K(I) \urcorner \sqsupseteq \ulcorner k^{-1} \urcorner$. That means, to alter this zone (to decrease the security level of $\alpha$), the intruder should have derived the atomic key $k^{-1}$ in advance. So, in this stage of the proof, his knowledge should satisfy the condition $\ulcorner K(I) \urcorner \sqsupseteq \ulcorner k^{-1} \urcorner$. Since the key $ k^{-1}$ satisfies the condition $\ulcorner k^{-1} \urcorner \sqsupseteq \ulcorner \alpha \urcorner$ then the knowledge of the intruder should satisfy the condition $\ulcorner K(I) \urcorner \sqsupseteq \ulcorner \alpha \urcorner$ too by transitivity of the order"$\sqsupseteq$" in a lattice. This is accurately the definition of a full-invariant-by-intruder function. Furthermore, these functions are  also well-formed by construction. Then, they are reliable. 

\begin{exemple}
Let $\alpha$ be an atom, $m$ be a message and $k_{ab}$ be a key such that:
$\ulcorner \alpha \urcorner=\{A, B, S\}$; $m=\{A.C.\{\alpha.D\}_{k_{as}}\}_{k_{ab}}$; ${k_{ab}^{-1}}={k_{ab}}, {k_{as}^{-1}}={k_{as}}$; $\ulcorner{k_{as}}\urcorner=\{A, S\}, \ulcorner{k_{ab}}\urcorner=\{A, B\}$;\\ 
$ $\\
Under the perfect  encryption assumption (empty equational theory), we have: \\
$S_{MAX}^{EK}(\alpha,m)=S_{MAX}^{EK}(\alpha,\{A.C.\{\alpha.D\}_{k_{as}}\}_{k_{ab}})=\{A, C, D, {k_{ab}^{-1}}\}$;\\ $F_{MAX}^{EK}(\alpha,m)=\psi\circ S_{MAX}^{EK}(\alpha,m)=\{A, C, D\}{ \sqcap}\ulcorner{k_{ab}^{-1}}\urcorner=\{A,C,D\} \cup \{A,B\}=\{A, C, D, B\}$.\\
$ $\\
Under the cipher homomorphism, we have: \\
$S_{MAX}^{EK}(\alpha,m)=S_{MAX}^{EK}(\alpha,\{A.C.\{\alpha.D\}_{k_{as}}\}_{k_{ab}})=S_{MAX}^{EK}(\alpha,\{A\}_{k_{ab}}.\{C\}_{k_{ab}}.\{\{\alpha.D\}_{k_{as}}\}_{k_{ab}})=
S_{MAX}^{EK}(\alpha,\{A\}_{k_{ab}})\cup S_{MAX}^{EK}(\alpha,\{C\}_{k_{ab}})\cup S_{MAX}^{EK}(\alpha,\{\{\alpha\}_{k_{as}}\}_{k_{ab}})\cup S_{MAX}^{EK}(\alpha,\{\{D\}_{k_{as}}\}_{k_{ab}})=\emptyset \cup \emptyset \cup \{{k_{ab}^{-1}}\}\cup \emptyset=\{{k_{ab}^{-1}}\}$; \\
$ $\\
$F_{MAX}^{EK}(\alpha,m)=\psi\circ S_{MAX}^{EK}(\alpha,m)=\ulcorner{k_{ab}^{-1}}\urcorner=\{A,B\}$.
\end{exemple}

In the rest of this paper, we denote by $F$ any of the functions $F_{MAX}^{EK}, F_{EK}^{EK}$ and $F_{N}^{EK}$.

\section{The witness-functions}\label{sectionWF}

From Theorem \ref{mainTh}, if a protocol $p$ is confirmed $F$-increasing on its \textit{valid traces} using a reliable function $F$, then it is correct with respect to secrecy. However, the set of traces is not finite. In order to be able to analyze a protocol on its finite set of the generalized roles, we have to readjust the reliable function so that it can deal with the problem of substitution and we seek an extra mechanism that enables us to pass from the decision made on  the generalized roles to the same decision on the ground terms of the valid traces. The witness-functions are designed for that purpose. But first, let us instill the notion of derivative messages. A derivative message is a term in the generalized roles from which we rule out the variables. This is described by Definition~\ref{derivation}. 

\begin{defini}{(Derivation)}\label{derivation}
{
A derivative message is defined as follows:
\begin{center}
\begin{tabular}{rrcll}
~~~~~~& $\partial_X \alpha$ & $=$ & $\alpha$ &\\
& $\partial_X \epsilon$ & $=$ & $\epsilon$ &\\
& $\partial_X X$ & $=$& $\epsilon$ &\\
& $\partial_X Y$ & $=$ & $Y$ &\\
& $\partial_{\{X\}} m$ & $=$ & $\partial_{X} m$ &\\
& $\partial{[\overline{X}]} m$ & $=$ & $\partial_{\{{\mathcal{X}}_m\backslash X\}} m$ &\\
& $\partial_X f(m)$ & $=$ & $ {f}(\partial_X m), f\in \Sigma$ &\\
& $\partial_{S_1 \cup S_2}m$ & $=$& $\partial_{S_1}\partial_{S_2}m$&\\
\\
\end{tabular}
\end{center}
}
\end{defini}

The idea now is to apply a reliable function ${F}$ to derivative messages istead of the message itself. For an atom in the static part of a message (i.e. in $\partial m$), we compute its security with no respect to variables. Else, for any content coming by substitution of a variable $X$, it is computed as the variable itself treated as a constant block. This is motivated by the fact that if the security of the block substituting $X$ does not decrease, then the whole block (the global secret $X\sigma$) is never revealed and hence any sub-secret in it is never revealed. This is given by Definition \ref{Fder}. 
\begin{defini}\label{Fder}
{
Let $m\in {\mathcal{M}}_p^{\mathcal{G}}$, $X \in {\mathcal{X}}_m$ and $m\sigma$ be a valid trace. 
For all $\alpha \in {\mathcal{A}}(m\sigma)$, $\sigma\in\Gamma$, we denote by:
\[
{F}(\alpha, \partial [\overline{\alpha}] m\sigma) = \left\{
\begin{array}{ll}
{F}(\alpha,\partial m) & \mbox{if } \alpha \in {\mathcal{A}}(\partial m),\\
{F}(X,\partial [\overline{X}] m) & \mbox{if }\alpha \notin {\mathcal{A}}(\partial m) \\
& \mbox{and } \alpha \in {\mathcal{A}}(X\sigma).
\end{array}
\right.
\]
}
\end{defini}

The application in Definition \ref{Fder} could not still be used to analyze protocols since derivation has a serious undesirable side-effect. Let have a look at Example \ref{exempleApp}:
\begin{exemple}\label{exempleApp}
{
Let $m_1$ and $m_2$ be two messages of ${\mathcal{M}}_p^{\mathcal{G}}$ such that 
$m_1=\{X.\alpha.D\}_{k_{ab}}$ and $m_2=\{C.\alpha.Y\}_{k_{ab}}$ and $\ulcorner\alpha\urcorner=\{A,B\}$. Let $m=\{C.\alpha.D\}_{k_{ab}}$ be in a valid trace.\\
$$F_{MAX}^{EK}(\alpha,\partial [\overline{\alpha}] m)=\begin{cases}
\{A, B, D\},& \!\!\!\mbox{if } m=m_1\sigma_1|X\sigma_1=C,\\
\{A, B, C\},& \!\!\!\mbox{if } m=m_2\sigma_2|Y\sigma_2=D
\end{cases}$$
Thus, $F_{MAX}^{EK}(\alpha,\partial [\overline{\alpha}] m)$ is not even a function. (i.e. it may return more than one value to the same input). 
}
\end{exemple}

The witness-function in Definition \ref{WF} fixes this bug: it looks for all the sources of $m\sigma$, applies the application in Definition \ref{Fder} and returns the minimum. This  minimum must exist and is unique in a lattice. 

\begin{defini}{(Witness-Function)}\label{WF}
{
Let $m\in {\mathcal{M}}_p^{\mathcal{G}}$, $X \in {\mathcal{X}}_m$ and $m\sigma$ be a valid trace.
Let $p$ be a protocol and $F$ be a ${\mathcal{C}}$-reliable Function.
We define a witness-function ${\cal{W}}_{p,{F}}$ for all $\alpha \in {\mathcal{A}}(m\sigma)$, $\sigma\in\Gamma$, as follows: 
\[{{{\cal{W}}}}_{p,{F}}(\alpha,m\sigma)=\!\!\!\!\!\!\!\!\!\!\underset{\overset {m' \in {\mathcal{M}}_p^{\mathcal{G}}}{\exists \sigma' \in \Gamma.m'\sigma' = m \sigma }}{\sqcap}\!\!\!\!\!\!\!\!\!\! {F}(\alpha, \partial [\overline{\alpha}] m'\sigma')\]
}
\end{defini}

A witness-function ${{{\cal{W}}}}_{p,{F}}$ is reliable when $F$ is reliable. In fact, it is easy to see that it is well-formed. It is also full-invariant-by-intruder as the returned values (principal identities) are those returned by $F$ on derivative messages of the sources of $m\sigma$ and derivation does not add new candidates, it just takes away some of them (that come by substitution), but returns always elements from the same invariant area in the message.

Since the goal of the witness-functions is to run a static analysis of the protocol and since it still depends on the protocol runs $\sigma$, we are going to confine the witness-functions in two static bounds that we will use for analysis instead of the witness-function itself. Proposition \ref{PropBoun} gives these bounds.

\begin{pre}{(Witness-Function Bounds)}\label{PropBoun}
{
Let $m\in{\mathcal{M}}_p^{\mathcal{G}}$.
Let $F$ be a ${\mathcal{C}}$-reliable function and ${\cal{W}}_{p,{F}}$ be a witness-function.
For all $\sigma \in \Gamma$ we have:
$${F}(\alpha,\partial[\overline{\alpha}] m)\sqsupseteq {{{\cal{W}}}}_{p,{F}}(\alpha,m\sigma)\sqsupseteq \!\!\!\!\!\!\!\!\!\!\!\! \underset{\overset{m' \in {\mathcal{M}}_p^{\mathcal{G}}}{\exists \sigma' \in \Gamma.m'\sigma' = m \sigma' }}{\sqcap} \!\!\!\!\!\!\!\!\!\!\!\!\!{F}(\alpha, \partial [\overline{\alpha}]m'\sigma')$$
}
\end{pre}

For a secret $\alpha$ in a ground term $m\sigma$, the upper-bound ${F}(\alpha,\partial[\overline{\alpha}] m)$ computes its security from one trivial source $m$ in the generalized roles. The witness-function ${{{\cal{W}}}}_{p,{F}}(\alpha,m\sigma)$ computes it from the set of the exact sources of $m\sigma$ where $m$ is necessarily one of them. The lower-bound $\!\!\!\!\!\!\!\!\!\! \underset{\overset{m' \in {\mathcal{M}}_p^{\mathcal{G}}}{\exists \sigma' \in \Gamma.m'\sigma' = m \sigma' }}{\sqcap} \!\!\!\!\!\!\!\!\!\!\!\!\!{F}(\alpha, \partial [\overline{\alpha}]m'\sigma')$ computes it from all the messages that could unify with $m$. This set necessarily includes the set of definition of the witness-function because the set of messages that unify with the ground term $m\sigma$ (fixed $\sigma$) is always in the set of messages that unify with $m$. Unifications in the lower-bound catch any odd principal identity inserted by the intruder. Please notice that the upper-bound and the lower-bound do not depend on $\sigma$ and are statically computable. Theorem~\ref{PAT} provides a static criterion for secrecy using these bounds. It is a direct result of Theorem~\ref{mainTh} and Proposition~\ref{PropBoun}. This enables a static analysis of the protocol to be run on the generalized roles and the decision to be  extended to valid traces. 

\begin{thm}{(Correctness Criterion)}\label{PAT}
{
Let $p$ be a protocol. 
Le $F$ be a reliable function.
Let ${\cal{W}}_{p,{F}}$ be a witness-function.
A sufficient condition for $p$ to be correct respect to secrecy is:\\
$\forall R.r \in R_G(p), \forall \alpha \in {\mathcal{A}}{(r^+ )}$ we have:
$$\underset{\overset {m' \in {\mathcal{M}}_p^{\mathcal{G}}}{\exists \sigma' \in \Gamma.m'\sigma' = r^+ \sigma' }}{\sqcap} \!\!\!\!\!\!\!\!\!\! {F}(\alpha, \partial[\overline{\alpha}] m'\sigma') \sqsupseteq \ulcorner \alpha \urcorner \sqcap {F}(\alpha,\partial[\overline{\alpha}] R^-)$$
}
\end{thm}


\section{Needham-Schroeder-Lowe protocol analysis with a witness-function under cipher homomorphism}\label{sectionAn1}

Hereafter, we analyze a variant of the Needham-Schroeder-Lowe protocol given in Table \ref{NSLVar} with a witness-function. 

\begin{table}

\begin{center}
\begin{tabular}{|llll|}
\hline 
&&& \\
  $\langle 1, A$ & $\longrightarrow$ & $B:$ & $\{N_a.A\}_{k_b}\rangle,$ \\
  $\langle 2,B$ & $\longrightarrow$ & $A:$ & $\{B.N_a\}_{k_a}.\{B.N_b\}_{k_a}\rangle,$ \\
  $\langle 3,A$ & $\longrightarrow$ & $B:$ & $A.B.\{N_b\}_{k_b}\rangle.$\\
&&&\\
\hline 
\end{tabular}
\end{center}
\caption{A variation of the Needham-Schroeder-Lowe protocol}\label{NSLVar}

\end{table}

The generalized roles of this protocol  in a role-based specification are $\cal{R}_{\cal{G}}(\textit{p}_{\textit{NSL}})$$=$ $\{A_{\cal{G}}^{\textit{1}},A_{\cal{G}}^{\textit{2}},B_{\cal{G}}^{\textit{1}},B_{\cal{G}}^{\textit{2}}\}$ where:

\begin{center}
\begin{tabular}{lllllll}
  $A_{\cal{G}}^{\textit{1}}=$ & $\alpha.1$ & $A$& $\longrightarrow I(B)$&:& $\{N_a^{\alpha}.A\}_{k_b} $ \\
  $A_{\cal{G}}^{\textit{2}}=$ & $\alpha.1$ & $A$& $\longrightarrow I(B)$&:& $\{N_a^{\alpha}.A\}_{k_b} $ \\
  $$ & $\alpha.2$ & $I(B)$&$ \longrightarrow A$ & : & $\{B.N_a^{\alpha}\}_{k_a}.\{B.X\}_{k_a}$ \\
  $$ & $\alpha.3$ & $A$& $\longrightarrow I(B)$ & : & $A.B.\{X\}_{k_b}$ \\
  \\
  $B_{\cal{G}}^{\textit{1}}=$ & $\alpha.1$ & $I(A)$&$ \longrightarrow B$ & : & $\{Y.A\}_{k_b}$\\
  $$ & $\alpha.2$ & $B$& $\longrightarrow I(A) $ & : & $\{B.Y\}_{k_a}.\{B.N_b^{\alpha}\}_{k_a}$ \\
  $B_{\cal{G}}^{\textit{2}}=$ & $\alpha.1$ & $I(A)$& $\longrightarrow B$ & : & $\{Y.A\}_{k_b}  $\\
  $$ & $\alpha.2$ & $B$&$ \longrightarrow I(A)$ & : & $\{B.Y\}_{k_a}. \{B.N_b^{\alpha}\}_{k_a}$ \\
  $$ & $\alpha.3$ & $I(A)$&$ \longrightarrow B$ & : & $A.B.\{N_b^{\alpha}\}_{k_b}$ 
\end{tabular}
\end{center}

$ $\\
Let us have a context of verification such that: $\ulcorner A\urcorner= \bot$; $\ulcorner B\urcorner= \bot$; $\ulcorner N_a^{\alpha}\urcorner= \{A,B\}$ (secret shared between $A$ and $B$); $\ulcorner N_b^{\alpha}\urcorner= \{A,B\}$ (secret shared between $A$ and $B$); $\ulcorner k_a^{-1}\urcorner=\{A\}$; $\ulcorner k_b^{-1}\urcorner=\{B\}$; $({\cal{L}},\sqsupseteq, \sqcup, \sqcap, \bot,\top)=(2^{\cal{I}},\subseteq,\cap,\cup,\cal{I}, \emptyset)$;
${\cal{I}}=\{I, A, B, A_1, A_2, $$B_1, B_2,...\}$;
\\
The set of messages generated by the protocol is ${\cal{M}}_p^{\cal{G}}=$
$\{\{N_{A_1}.A_1\}_{k_{B_1}},
\{B_2.N_{A_2}\}_{k_{A_2}},\\
\{B_3.X_1\}_{k_{A_3}},
\{X_2\}_{k_{B_4}},
\{Y_1.A_{4}\}_{k_{B_5}},
\{B_6.Y_2\}_{k_{A_5}},\\
\{B_7.N_{B_7}\}_{k_{A_6}},
\{N_{B_8}\}_{k_{B_8}}
 \}
 $\\
The variables are denoted by $X_1, X_2, Y_1$ and $Y_2$;\\ 
The static names are denoted by $N_{A_1}$, $A_1$, $k_{B_1}$, $B_2$, $N_{A_2}$, $k_{A_2}$, $B_3$, $k_{A_3}$, $k_{B_4}$, $A_{4}$, $k_{B_5}$, $B_6$, $k_{A_5}$, $B_7$, $N_{B_7}$, $k_{A_6}$, $N_{B_8}$ and $k_{B_8}$.\\
\\
After duplicates removal, ${\cal{M}}_p^{\cal{G}}=$
$\{\{N_{A_1}.A_1\}_{k_{B_1}},
\{B_2.N_{A_2}\}_{k_{A_2}},
\{B_3.X_1\}_{k_{A_3}},
\{X_2\}_{k_{B_4}},\\
\{Y_1.A_{4}\}_{k_{B_5}},
\{B_7.N_{B_7}\}_{k_{A_6}},
\{N_{B_8}\}_{k_{B_8}}
 \}
 $\\
Let us define the witness-function as follows:\\
$p$ = Needham-Schroeder-Lowe; $F=F_{MAX}^{EK}$; ${\cal{W}}_{p,F}(\alpha,m\sigma)=\underset{\overset {m' \in {\cal{M}}_p^{\cal{G}}}{\exists \sigma'.m'\sigma' = m\sigma }}{\sqcap}\!\!\!\!\!\!\!\!\!\! F(\alpha, \partial[\overline{\alpha}] m'\sigma')$;\\
\\
Let us denote the lower-bound of the witness-function in Theorem \ref{PAT} by:$${\cal{W}}'_{p,F}(\alpha,r^+)=\underset{\overset {m' \in {\cal{M}}_p^{\cal{G}}}{\exists \sigma' \in \Gamma.m'\sigma'=r^+\sigma'}}{\sqcap}\!\!\!\!\!\!\!\!\!\! F(\alpha, \partial [\overline{\alpha}]m'\sigma')$$ 
The principal identities in the context are not analyzed since they are public.\\ 
$ $\\
The protocol is analyzed under cipher homomorphism (i.e. $\{m.m'\}_k=\{m\}_k.\{m'\}_k$). The smallest selection for any $\alpha\in {\mathcal{A}(m.m')}$ is in the form $\{m\}_k.\{m'\}_k$.\\ The protocol is analyzed for secrecy only.\\ It is important to recall that this protocol has been established correct for secrecy under the perfect encryption assumption using the witness-functions\cite{WitnessArt1} and by many other techniques.
\subsection{Analysis of the generalized role of $A$}
As defined in the generalized roles of $p$, an agent $A$ can participate in two subsequent sessions: $S^{i}$ and $S^{j}$ such that $j>i$. In the former session $S^{i}$, the agent $A$ receives nothing and sends the message $\{N_a^{\alpha}.A\}_{k_b} $. In the subsequent session $S^{j}$, he receives the message $\{B.N_a^{\alpha}\}_{k_a}.\{B.X\}_{k_a}$ and he sends the message $A.B.\{X\}_{k_b}$. This is described by the following rules:
\[{S^{i}}:\frac{\Box}{\{N_a^{\alpha}.A\}_{k_b} } ~~~~~~~~~~~~~~~~~~~~~~~~ {S^{j}}:\frac{\{B.N_a^{\alpha}\}_{k_a}.\{B.X\}_{k_a}}{A.B.\{X\}_{k_b}}\]
\textbf{Analysis of the messages exchanged  in the session $S^{i}$:}\\
\\
1- For $N_a^{\alpha}$:\\
a- On sending: $r_{S^{i}}^+=\{N_a^{\alpha}.A\}_{k_b}$ \textit{(in a sending step, the lower-bound is used)}\\
 $N_a^{\alpha}.\{m' \in {\cal{M}}_p^{\cal{G}}|{\exists \sigma'\in\Gamma.\sigma'm'=\sigma'r_{S^{i}}^+) } \}$\\$= N_a^{\alpha}.\{m' \in {\cal{M}}_p^{\cal{G}}|{\exists \sigma'\in\Gamma.m'\sigma'=\{N_a^{\alpha}.A\}_{k_b}\sigma') } \}$ \\$=\{(\{N_{A_1}.A_1\}_{k_{B_1}},\sigma_1'), (\{X_2\}_{k_{B_4}},\sigma_2'), (\{Y_1.A_{4}\}_{k_{B_5}},\sigma_3')\}$ such that: \\
$$\left\{
    \begin{array}{l}
      \sigma_1'=\{N_{A_1} \longmapsto N_a^{\alpha}, A_1 \longmapsto A, k_{B_1} \longmapsto k_b\}   \\
      \sigma_2'=\{ X_2 \longmapsto N_a^{\alpha}.A, k_{B_4} \longmapsto k_b\}  \\
      \sigma_3'=\{Y_1 \longmapsto N_a^{\alpha},  A_4 \longmapsto A,  k_{B_5} \longmapsto {k_b} \}
    \end{array}
\right.$$
${\cal{W}}_{p,F}'(N_a^{\alpha},\{N_a^{\alpha}.A\}_{k_b})$\\
$=\{\mbox{Definition of the lower-bound of the witness-function}\}$\\
$F(N_a^{\alpha},\partial[\overline{N_a^{\alpha}}]\{N_{A_1}.A_1\}_{k_{B_1}} \sigma_{1}') \sqcap F(N_a^{\alpha},\partial[\overline{N_a^{\alpha}}] \{X_2\}_{k_{B_4}} \sigma_{2}') \sqcap$ $F(N_a^{\alpha},\partial [\overline{N_a^{\alpha}}]\{Y_1.A_{4}\}_{k_{B_5}}\sigma_{3}')$\\
$=\{\mbox{Setting the static neighborhood}\}$\\
$F(N_a^{\alpha},\partial[\overline{N_a^{\alpha}}]\{{N_a^{\alpha}}.A\}_{k_{b}} \sigma_{1}') \sqcap F(N_a^{\alpha},\partial[\overline{N_a^{\alpha}}] \{X_2\}_{k_{b}} \sigma_{2}') \sqcap$  $F(N_a^{\alpha},\partial [\overline{N_a^{\alpha}}]\{Y_1.A\}_{k_{b}}\sigma_{3}')$\\
$=\{\mbox{Definition } \ref{Fder}\}$
\\
$F(N_a^{\alpha},\{{N_a^{\alpha}}.A\}_{k_{b}}) \sqcap F({X_2},\partial[\overline{X_2}] \{X_2\}_{k_{b}} ) \sqcap$  $F(Y_1,\partial [\overline{Y_1}]\{Y_1.A\}_{k_{b}})$\\
$=\{\mbox{Definition } \ref{derivation}\}$
\\
$F(N_a^{\alpha},\{N_a^{\alpha}.A\}_{k_b}) \sqcap F(X_2,\{X_2\}_{k_b}) \sqcap F(Y_1, \{Y_1.A\}_{k_b})$\\
$=\{\mbox{Normal form under homomorphism}\}$
\\
$F(N_a^{\alpha},\{N_a^{\alpha}\}_{k_b}) \sqcap F(N_a^{\alpha},\{A\}_{k_b}) \sqcap F(X_2,\{X_2\}_{k_b}) \sqcap F(Y_1, \{Y_1\}_{k_b})\sqcap F(Y_1, \{A\}_{k_b})$\\
$=\{\mbox{Since } F=F_{MAX}^{EK}\}$\\
$\{B\} \sqcap \top \sqcap \{B\} \sqcap \{B\} \sqcap \top=\{B\}$(1.0)\\
\\
b- On receiving: $R_{S^{i}}^-=\emptyset$ \textit{(in a receiving step, the upper-bound is used)}\\
$F(N_a^{\alpha},\partial  [\overline{N_a^{\alpha}}]\emptyset )=F(N_a^{\alpha},\emptyset)=\top$ (1.1)\\
\\
2- Conformity with Theorem \ref{PAT}:\\
From (1.0) and (1.1), we have: ${\cal{W}}_{p,F}'(N_a^{\alpha},\{N_a^{\alpha}.A\}_{k_b})= \{B\} \sqsupseteq \ulcorner N_a^{\alpha} \urcorner \sqcap F(N_a^{\alpha},\partial  [\overline{N_a^{\alpha}}]\emptyset )= \ulcorner N_a^{\alpha} \urcorner \sqcap \top=\{A,B\}$ (1.2)\\
From (1.2) we have: the messages exchanged in the session $S^{i}$ respect the correctness criterion set in Theorem \ref{PAT}. (I)\\
$ $\\
\textbf{Analysis of the messages exchanged  in the session $S^{j}$:}\\
\\
1-$\forall X$:\\
a- On sending: $r_{S^{j}}^+=A.B.\{X\}_{k_b}$\\
${\cal{W}}_{p,F}'(X,A.B.\{X\}_{k_b})={\cal{W}}_{p,F}'(X,A) \sqcap {\cal{W}}_{p,F}'(X,B) \sqcap {\cal{W}}_{p,F}'(X,\{X\}_{k_b})=\top \sqcap \top \sqcap {\cal{W}}_{p,F}'(X,\{X\}_{k_b})={\cal{W}}_{p,F}'(X,\{X\}_{k_b})$ (2.0)\\
$\forall X.\{m' \in {\cal{M}}_p^{\cal{G}}|{\exists \sigma'\in\Gamma.m'\sigma'=\{X\}_{k_b}\sigma') } \}$
$=\{ (\{X_2\}_{k_{B_4}},\sigma_1') \}$ such that:
$$\sigma'_1=\{X_2 \longmapsto X, k_{B_4} \longmapsto k_b \}$$
${\cal{W}}_{p,F}'(X,\{X\}_{k_b})$\\
$=\{\mbox{Definition of the lower-bound of the witness-function}\}$\\
$F(X,\partial [\overline{X}]\{X_2\}_{k_{B_4}}\sigma'_{1})$\\
$=\{\mbox{Setting the static neighborhood}\}$\\
$F(X,\partial [\overline{X}]\{X_2\}_{k_{b}}\sigma'_{1})$\\
$=\{\mbox{Definition } \ref{Fder}\}$\\
$F(X_2,\partial [\overline{X_2}]\{X_2\}_{k_{b}})$\\
$=\{\mbox{Definition } \ref{derivation}\}$\\
$F(X_2, \{X_2\}_{k_{b}})$\\
$=\{\mbox{Normal form under homomorphism}\}$
\\
$F(X_2, \{X_2\}_{k_{b}})$\\
$=\{\mbox{Since } F=F_{MAX}^{EK}\}$\\
$\{B\}$ (2.1)\\
\\
b- On receiving: $R_{S^{j}}^-=\{B.N_a^{\alpha}\}_{k_a}.\{B.X\}_{k_a}$ \textit{(in a receiving step, the upper-bound is used)}\\
$F(X,\partial[\overline{X}]\{B.N_a^{\alpha}\}_{k_a}.\{B.X\}_{k_a})=$$F(X,\partial[\overline{X}]\{B.N_a^{\alpha}\}_{k_a})\sqcap F(X,\partial[\overline{X}]\{B.X\}_{k_a})=$\\$F(X,\{B.N_a^{\alpha}\}_{k_a})\sqcap F(X,\{B.X\}_{k_a})$\\
$=\{\mbox{Normal form under  homomorphism}\}$\\
$F(X,\{B\}_{k_a})\sqcap F(X,\{N_a^{\alpha}\}_{k_a})\sqcap F(X,\{B\}_{k_a})\sqcap F(X,\{X\}_{k_a})=$\\
$\top \sqcap \top \sqcap \top \sqcap \{A\}=\{A\}$
(2.2)\\
\\
3-Conformity with Theorem \ref{PAT}:\\
From (2.0), (2.1) and (2.2), we have:\\
${\cal{W}}_{p,F}'(X,A.B.\{X\}_{k_b})=\{B\}  \not \sqsupseteq \ulcorner X \urcorner \sqcap F(X,\partial[\overline{X}]\{B.N_a^{\alpha}\}_{k_a}.\{B.X\}_{k_a})= \ulcorner X \urcorner \cup \{A\}$ (2.3)\\
From (2.3), we have: the messages exchanged in the session $S^{j}$ do not respect the correctness criterion set in Theorem \ref{PAT}. (II)\\
From (I) and (II), the messages exchanged in the generalized role of $A$ do not respect the correctness criterion set in Theorem \ref{PAT}. (III)

\subsection{Analysis of the generalized role of $B$}
As defined in the generalized roles of $p$, an agent $B$ can participate in a session ${{S'}^{^{i}}}$, in which he receives the message $\{Y.A\}_{k_b}$ and he sends the message $\{B.Y\}_{k_a}.\{B.N_b^{\alpha}\}_{k_a}$. This is described by the following rule:
\[{{S'}^{^{i}}}:\frac{\{Y.A\}_{k_b}}{\{B.Y\}_{k_a}.\{B.N_b^{\alpha}\}_{k_a}}
\]
1- For $N_b^{\alpha}$:\\
a- On sending: $r_{{S'}^{^{i}}}^+=\{B.Y\}_{k_a}.\{B.N_b^{\alpha}\}_{k_a}$   \textit{(in a sending step, the lower-bound is used)}\\
${\cal{W}}_{p,F}'(N_b^{\alpha},\{B.Y\}_{k_a}.\{B.N_b^{\alpha}\}_{k_a})=$${\cal{W}}_{p,F}'(N_b^{\alpha},\{B.Y\}_{k_a}) \sqcap {\cal{W}}_{p,F}'(N_b^{\alpha},\{B.N_b^{\alpha}\}_{k_a})=$\\$\top \sqcap {\cal{W}}_{p,F}'(N_b^{\alpha},\{B.N_b^{\alpha}\}_{k_a})={\cal{W}}_{p,F}'(N_b^{\alpha},\{B.N_b^{\alpha}\}_{k_a})$ (3.0)\\
$\forall N_b^{\alpha}.\{m' \in {\cal{M}}_p^{\cal{G}}|{\exists \sigma'\in \Gamma.m'\sigma'=\{B.N_b^{\alpha}\}_{k_a}\sigma') }\}$\\
$=\{(\{B_3.X_1\}_{k_{A_3}},\sigma'_1), (\{X_2\}_{k_{B_4}},\sigma'_2), (\{B_7.N_{B_7}\}_{k_{A_6}},\sigma'_3) \}$ such that:
$$
\left\{
    \begin{array}{l}
     \sigma'_1=\{B_3 \longmapsto B, X_1 \longmapsto N_b^{\alpha} ,  k_{A_3} \longmapsto k_a \}    \\
      \sigma'_2=\{X_2 \longmapsto B.N_b^{\alpha}, k_{B_4} \longmapsto k_a \}  \\
      \sigma'_3=\{B_7 \longmapsto B, N_{B_7} \longmapsto N_b^{\alpha} , k_{A_6} \longmapsto k_a\}
    \end{array}
\right.
 $$
${\cal{W}}_{p,F}'(N_b^{\alpha},\{B.N_b^{\alpha}\}_{k_a})$\\
$=\{\mbox{Definition of the lower-bound of the witness-function}\}$\\
$F(N_b^{\alpha},\partial [\overline{N_b^{\alpha}}] \{B_3.X_1\}_{k_{A_3}}\sigma'_{1}) \sqcap F(N_b^{\alpha},\partial [\overline{N_b^{\alpha}}] \{X_2\}_{k_{B_4}}\sigma'_{2})\sqcap $$ F(N_b^{\alpha},\partial[\overline{N_b^{\alpha}}] \{B_7.N_{B_7}\}_{k_{A_6}}\sigma'_{3})$\\
$=\{\mbox{Setting the static neighborhood}\}$\\
$F(N_b^{\alpha},\partial [\overline{N_b^{\alpha}}] \{B.X_1\}_{k_{a}}\sigma'_{1}) \sqcap F(N_b^{\alpha},\partial [\overline{N_b^{\alpha}}] \{X_2\}_{k_{a}}\sigma'_{2})\sqcap $$ F(N_b^{\alpha},\partial[\overline{N_b^{\alpha}}] \{B.N_b^{\alpha}\}_{k_{a}}\sigma'_{3})$\\
$=\{\mbox{Definition } \ref{Fder}\}$\\
$F(X_1,\partial [\overline{X_1}] \{B.X_1\}_{k_{a}}) \sqcap F(X_2,\partial [\overline{X_2}] \{X_2\}_{k_{a}})$ $\sqcap $  
$ F(N_b^{\alpha}, \partial [\overline{N_b^{\alpha}}]\{B.N_b^{\alpha}\}_{k_{a}})$\\
$=\{\mbox{Definition } \ref{derivation}\}$\\
$F(X_1,\{B.X_1\}_{k_a}) \sqcap F(X_2,\{X_2\}_{k_a}) \sqcap F(N_b^{\alpha},\{B.N_b^{\alpha}\}_{k_a})$\\
$=\{\mbox{Normal form under homomorphism}\}$
\\
$F(X_1,\{B\}_{k_a}) \sqcap F(X_1,\{X_1\}_{k_a})  \sqcap F(X_2,\{X_2\}_{k_a}) \sqcap F(N_b^{\alpha},\{B\}_{k_a})\sqcap F(N_b^{\alpha},\{N_b^{\alpha}\}_{k_a})$\\
$=\{\mbox{Since } F=F_{MAX}^{EK}\}$\\
$\top\sqcap\{A\}\sqcap\{A\}\sqcap \{A\} \sqcap \top \sqcap\{A\}=\{A\}$ (3.1)\\
\\
b- On receiving: $R_{{S'}^{^{i}}}^-=\{Y.A\}_{k_b}$  \textit{(in a receiving step, the upper-bound is used)}\\
$F(N_b^{\alpha},\partial [\overline{N_b^{\alpha}}]\{Y.A\}_{k_b})=F(N_b^{\alpha},\{A\}_{k_b})=\top$ (3.2)\\
\\
2- $\forall Y$:\\
a- On sending: $r_{{S'}^{^{i}}}^+=\{B.Y\}_{k_a}.\{B.N_b^{\alpha}\}_{k_a}$ \textit{(in a receiving step, the upper-bound is used)}\\
${\cal{W}}_{p,F}'(Y,\{B.Y\}_{k_a}.\{B.N_b^{\alpha}\}_{k_a})={\cal{W}}_{p,F}'(Y,\{B.Y\}_{k_a}) \sqcap {\cal{W}}_{p,F}'(Y,\{B.N_b^{\alpha}\}_{k_a})=$\\${\cal{W}}_{p,F}'(Y,\{B.Y\}_{k_a}) \sqcap \top ={\cal{W}}_{p,F}'(Y,\{B.Y\}_{k_a})$ (3.3)\\
\\
$\forall Y.\{m' \in {\cal{M}}_p^{\cal{G}}|{\exists \sigma'\in\Gamma.m'\sigma'=\{B.Y\}_{k_a}\sigma') }\}$
$=\{(\{B_3.X_1\}_{k_{A_3}},\sigma_1), (\{X_2\}_{k_{B_4}},\sigma_2)\}$ such that:
$$
 \left\{
    \begin{array}{l}
     \sigma'_1=\{B_3 \longmapsto B, X_1 \longmapsto Y ,  k_{A_3} \longmapsto k_a \}    \\
      \sigma'_2=\{X_2 \longmapsto B.Y, k_{B_4} \longmapsto k_a \}
    \end{array}
\right.
$$
${\cal{W}}_{p,F}'(Y,\{B.Y\}_{k_a})$\\
$=\{\mbox{Definition of the lower-bound of the witness-function}\}$\\
$F(Y,\partial[\overline{Y}] \{B_3.X_1\}_{k_{A_3}}\sigma'_{1})$
$ \sqcap F(Y,\partial [\overline{Y}]\{X_2\}_{k_{B_4}}\sigma'_{2})$\\
$=\{\mbox{Setting the static neighborhood}\}$\\
$F(Y,\partial[\overline{Y}] \{B.X_1\}_{k_{a}}\sigma'_{1})$
$ \sqcap F(Y,\partial [\overline{Y}]\{X_2\}_{k_{a}}\sigma'_{2})=$\\
$=\{\mbox{Definition } \ref{Fder}\}$\\
$F(X_1,\partial[\overline{X_1}] \{B.X_1\}_{k_{a}})$
$ \sqcap F(X_2,\partial [\overline{X_2}]\{X_2\}_{k_{a}})$\\
$=\{\mbox{Definition } \ref{derivation}\}$\\
$F(X_1,\{B.X_1\}_{k_{a}}) \sqcap F(X_2,\{X_2\}_{k_{a}})$\\
$=\{\mbox{Normal form under homomorphism}\}$
\\
$F(X_1,\{B\}_{k_{a}}) \sqcap F(X_1,\{X_1\}_{k_{a}}) \sqcap F(X_2,\{X_2\}_{k_{a}})$\\
$=\{\mbox{Since } F=F_{MAX}^{EK}\}$\\
$ \top \sqcap \{A\} \sqcap \{A\}=\{A\}$ (3.4)\\
\\
b- On receiving: $R_{{S'}^{^{i}}}^-=\{Y.A\}_{k_b}$ \textit{(in a receiving step, the upper-bound is used)}\\
$F(Y,\partial [\overline{Y}]\{Y.A\}_{k_{b}})=F(Y,\{Y.A\}_{k_{b}})$
$=\{\mbox{Normal form under homomorphism}\}$
\\
$F(Y,\{Y\}_{k_{b}}) \sqcap F(Y,\{A\}_{k_{b}})$
$=\{B\}\sqcap\top=\{B\}$ (3.5)\\
\\
3- Conformity with Theorem \ref{PAT}:\\
From (3.0), (3.1) and (3.2) we have:\\
${\cal{W}}_{p,F}'(N_b^{\alpha},\{B.Y\}_{k_a}.\{B.N_b^{\alpha}\}_{k_a})=\{A\} \sqsupseteq \ulcorner N_b^{\alpha} \urcorner \sqcap F(N_b^{\alpha},\partial [\overline{N_b^{\alpha}}] \{Y.A\}_{k_b})=\{A,B\}$ (3.6)\\
From (3.3), (3.4) and (3.5) we have:\\
${\cal{W}}_{p,F}'(Y,\{B.Y\}_{k_a}.\{B.N_b^{\alpha}\}_{k_a})=\{A\} \not \sqsupseteq \ulcorner Y \urcorner \sqcap F(Y,\partial [\overline{Y}]\{Y.A\}_{k_{b}})= \ulcorner Y \urcorner \sqcap \{B\}$ (3.7)\\
From  (3.7), the messages exchanged in the session ${{S'}^{^{i}}}$ do not respect the correctness criterion set in Theorem \ref{PAT}. (IV)\\
From (IV), the messages exchanged in the generalized role of $B$ do not respect the correctness criterion stated Theorem \ref{PAT}. (V)

\subsection{Results and interpretation}
The results of the analysis of the Needham-Schroeder-Lowe protocol under homomorphism are summarized in Table \ref{NSLGrowth}.

\begin{table}[h]
   \centering
\scalebox{0.9}{
\begin{tabular}{|p{0.05cm}|p{0.3cm}|p{0.3cm}|p{2.45cm}|p{2.45cm}|p{1.1cm}|}
  \hline
  &$\alpha$ &  Role & $R^-$ & $r^+$ & Theo.\ref{PAT}\\
  \hline
  \hline
1 & $N_a^{\alpha}$ & $A$& $\emptyset$ & $\{N_a^{\alpha}.A\}_{k_b}$ & \ding{52}\\
  \hline
2& $\forall X$ &  $A$ & $\{B.N_a^{\alpha}\}_{k_a}.\{B.X\}_{k_a}$ & $A.B.\{X\}_{k_b}$ & \ding{54}\\
  \hline
 \hline
3& $\forall Y$ & $B$ & $\{A.Y\}_{k_b}$ & $\{B.Y\}_{k_a}. \{B.N_b^{\alpha}\}_{k_a}$ &\ding{54}\\
 \hline
4&  $N_b^{\alpha}$ & $B$ & $\{A.Y\}_{k_b}$ & $\{B.Y\}_{k_a}. \{B.N_b^{\alpha}\}_{k_a}$ & \ding{52}\\
   \hline
\end{tabular}
}
\caption[]{\label{NSLGrowth} Conformity of the Needham-Schroeder-Lowe protocol with Theorem \ref{PAT} under cipher homomorphism}
\end{table}

\begin{figure}[h]
\begin{mdframed}
   \centering

\begin{picture}(68,68)(68,-54)
\put(0,-3){A}
\put(10,8){$\{N^{\alpha}_{a}.A\}_{k_b}$}
\put(5,-1){
  \vector(4,0){53}
}
\put(62,-5){I}
\put(60,-6){
   \vector(0,-1){50}
}
\put(62,-65){B}
\put(67,-30){$\{N^{\alpha}_{a}.A\}_{k_b}.\{I\}_{k_b}=_{\xi}\{\underbrace{N^{\alpha}_{a}.A}_{N^{\beta}_{I}}.I\}_{k_b}$}
\put(68,-62){
  \vector(4,0){125}
}

\put(200,-65){I}
\put(90,-74){$\{B.\underbrace{N^{\alpha}_{a}.A}_{N^{\beta}_{I}}\}_{k_i}.\{B.N^{\beta}_{b}\}_{k_i}$}

\end{picture}
\\\hspace{\linewidth}
\\\hspace{\linewidth}
\\\hspace{\linewidth}
\end{mdframed}
\caption{Attack on the Needham-Schroeder-Lowe protocol under cipher homomorphism}
\label{FailleNSETH} 

\end{figure}
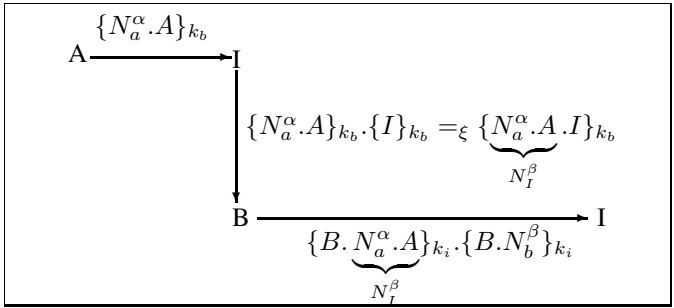

From  the rows (2) and (3) of Table \ref{NSLGrowth}, the Needham-Schroeder-Lowe protocol under the homomorphic property is rejected by  Theorem \ref{PAT}. Therefore, we conclude that it may involve a flaw with respect to secrecy. This flaw is described by Figure \ref{FailleNSETH}. In fact, an intruder can intercept the message $\{N^{\alpha}_{a}.A\}_{k_b}$ sent by a regular agent $B$ to a regular agent $A$. Then, he concatenates it to the message $\{I\}_{k_b}$ that he creates by himself. The intruder knows in advance that the resulting message $\{N^{\alpha}_{a}.A\}_{k_b}.\{I\}_{k_b}$ is equivalent to $\{N^{\alpha}_{a}.A.I\}_{k_b}$ under the homomorphic property. Then, he  initiates a new session with $B$ and sends him this resulting message. On recpetion, $B$ understands the string $N^{\alpha}_{a}.A$ in the received message as  a regular nonce $N^{\beta}_{I}$ from a regular agent $I$ starting a new session of the protocol since in a role-based specification this string corresponds to a variable $Y$ on which he cannot perform any verification. $B$ replies so by $\{B.N^{\alpha}_{a}.A\}_{k_i}.\{B.N^{\beta}_{b}\}_{k_i}$. The intruder has just to decrypt it to get  the secret $N^{\alpha}_{a}$ shared between $A$ and $B$ . The bounds of the used witness-function react well to this scenario and declares the drop of $Y$ in the generalized role of $B$. That is because they base their calculation on the static neighborhood only.  This neighborhood cannot be augmented by the intruder neither using his capabilities nor using the equational theory. The use of the normal form that gets rid of all the doubtful atoms is crucial for an analysis using the witness-functions under nonempty theories.

\subsection{Proposal of an amended  version}
To correct this variant of the Needham-Schroeder-Lowe protocol, we propose the amended version in Table \ref{NSLVarCorr}. In this version, the nonces $N_a$ and $N_b$ are sent back concatenated with the sender and hashed by a secure hash function, $\hbar ash$. The receiver has just to compare the hashed values with $\hbar ash(\mbox{sender}.\mbox{sent-nonce})$ to decide acceptance or rejection. Believing in the infeasiblility to generate a message from its digest, the nonces are never derived and the protocol keeps its secrects.

\begin{table}[h]

\begin{center}
\begin{tabular}{|llll|}
\hline 
&&& \\
  $\langle 1,A$ & $\longrightarrow$ & $B:$ & $\{N_a.A\}_{k_b}\rangle,$ \\
  $\langle 2,B$ & $\longrightarrow$ & $A:$ & $\{B.\hbar ash(B.N_a)\}_{k_a}.\{B.N_b\}_{k_a}\rangle,$ \\
  $\langle 3,A$ & $\longrightarrow$ & $B:$ & $A.\{\hbar ash(A.N_b)\}_{k_b}\rangle. $\\
&&& \\
\hline 
\end{tabular}
\end{center}
\caption{Amended  version of the Needham-Schroeder-Lowe protocol (proposal)}\label{NSLVarCorr}
\end{table}

\section{Related Works}

Under nonempty equational theories, our witness-functions 
could be compared to the interpretation-functions of Houmani~\cite{Houmani1,Houmani3,Houmani8,Houmani5}. Unfortunately, these functions often fail to describe flaws inside protocols and simply report the protocol unsecurity. They yield a high level of false negatives as well because they are not variable free in output. Contrariwise, the witness-functions are variable free in output owing to the derivation in its composition. 
We believe that our witness-functions are able to deal with other algebraic properties like the modular exponentiation property.


\section{Conclusion and Future Work}

In this paper, we presented how to use the witness-functions under nonempty equational theories to prove the correctness of cryptographic protocols with respect to secrecy. The major contribution is to adjut the witness-functions to deal with the algebraic properties in the equational theory through a judicious choice of the normal form on which we apply them. This normal form is obtained by a careful orietation of the rewriting system extracted from the theory. Afterwards, we successfully analyzed the Needham-Schroeder-Lowe protocol under the homomorphic encryption and we clearly provided an attack scenario on it. In a future work, we intend to analyze more protocols under different  theories~\cite{Pigozzi1979117,cortier9900,cortier1971,cortier9901}.

\section*{Notice}
\small{
© 2018 IEEE. Personal use of this material is permitted. Permission from IEEE must be obtained for all other uses, in any current or future media, including reprinting/republishing this material for advertising or promotional purposes, creating new collective works, for resale or redistribution to servers or lists, or reuse of any copyrighted component of this work in other works.
}

\bibliographystyle{unsrt}
\bibliographystyle{ieeetr} 
\bibliography{Ma_these} 

\normalsize

\end{document}